\documentclass[iop]{emulateapj}
\usepackage{color}
\usepackage{natbib}
\usepackage{amsmath}
\shorttitle{Radio Emission Studies of Dust-free Red Giants}
\shortauthors{O'Gorman et al.}

\begin{document}

\title{Multi-wavelength Radio Continuum Emission Studies of Dust-free Red Giants}

\author{Eamon O'Gorman\altaffilmark{1}, Graham M. Harper\altaffilmark{1}, Alexander Brown\altaffilmark{2}, Stephen Drake\altaffilmark{3}, and Anita M. S. Richards\altaffilmark{4}}
\altaffiltext{1}{School of Physics, Trinity College Dublin, Dublin 2, Ireland}
\altaffiltext{2}{Center for Astrophysics and Space Astronomy, University of Colorado, 389 UCB, Boulder, CO 80309, USA}
\altaffiltext{3}{NASA Goddard Space Flight Center, Greenbelt, MD 20771, USA}
\altaffiltext{4}{Jodrell Bank Centre for Astrophysics, School of Physics and Astronomy, University of Manchester, Manchester M13 9PL, UK}


\begin{abstract}

Multi-wavelength centimeter continuum observations of non-dusty, non-pulsating K spectral-type red giants directly sample their chromospheres and wind acceleration zones. Such stars are feeble emitters at these wavelengths however, and previous observations have provided only a small number of modest signal-to-noise measurements slowly accumulated over three decades. We present multi-wavelength Karl G. Jansky Very Large Array thermal continuum observations of the wind acceleration zones of two dust-free red giants, Arcturus ($\alpha$ Boo: K2 III) and Aldebaran ($\alpha$ Tau: K5 III). Importantly, most of our observations of each star were carried out over just a few days, so that we obtained a snapshot of the different stellar atmospheric layers sampled at different wavelengths, independent of any long-term variability. We report the first detections at several wavelengths for each star including a detection at 10 cm (3.0 GHz: \textit{S} band) for both stars and a 20 cm (1.5 GHz: \textit{L} band) detection for $\alpha$ Boo. This is the first time single (non-binary) luminosity class III red giants have been detected at these continuum wavelengths. Our long-wavelength data sample the outer layers of $\alpha$ Boo's atmosphere where its wind velocity is approaching (or possibly has reached) its terminal value and the ionization balance is becoming \textit{frozen-in}. For $\alpha$ Tau, however, our long-wavelength data are still sampling its inner atmosphere, where the wind is still accelerating probably due to its lower mass-loss rate. We compare our data with published semi-empirical models based on ultraviolet data, and the marked deviations highlight the need for new atmospheric models to be developed. Spectral indices are used to discuss the possible properties of the stellar atmospheres, and we find evidence for a rapidly cooling wind in the case of $\alpha$ Boo. Finally, we develop a simple analytical wind model for $\alpha$ Boo based on our new long-wavelength flux measurements.

\end{abstract}

\keywords{radio continuum: stars$-$stars: chromospheres$-$stars: individual ($\rm{\alpha}$ Boo, $\rm{\alpha}$ Tau)$-$stars: late-type$-$stars: winds, outflows}

\section{INTRODUCTION}

In order to understand the mechanisms which drive the $10^{-9}-10^{-11}$ $M_{\odot}$ yr$^{-1}$ mass-loss rates from evolved spectral-type K through mid-M stars, an understanding of the dynamics and thermodynamics of their atmospheres is essential. An important discovery in late-type evolved stellar atmospheres resulted from the first ultraviolet survey of such stars using the \textit{International Ultraviolet Explorer} (\textit{IUE}). The survey revealed a ``transition region dividing line'' in the giant branch near spectral type K1 III which separates these stars based on the properties of their atmospheres \citep{1979ApJ...229L..27L}. Stars blueward of the dividing line were found to possess chromospheres and transition regions like the Sun, while stars on the red side were found to possess chromospheres and cool winds. X-ray observations showed that this dividing line extended to coronal emission \citep{1981ApJ...250..293A}. Around the same time, another class of late-type evolved star emerged which showed signs of possessing both a transition region and a cool wind \citep[e.g.,][]{1982A&A...107..292R}. Many of these so-called hybrid atmosphere stars now also show evidence for coronal emission, albeit much weaker than on the blue side of the dividing line \citep{1997ApJ...491..876A,2005ApJ...622..629D}. Understanding the nature of the atmospheric structure of late-type evolved stars will ultimately lead to a broader understanding of the mass-loss process.

Mass-loss from late-type evolved stars plays a crucial role in both stellar and galactic evolution and ultimately provides part of the material required for the next generation of stars and planets. Despite the importance of this phenomenon and decades of study, the mechanisms that drive winds from evolved spectral-type K through mid-M stars remain an enduring mystery (clearly laid out by \citealt{1985ASSL..117..229H} but still unsolved, e.g., \citealt{2009AIPC.1094..267C}). There is insufficient atomic, molecular, or dust opacity to drive a radiation-driven outflow \citep{1995ApJ...446L..79Z,2008MNRAS.387..845J}, and acoustic/pulsation models cannot drive the observed mass-loss rates \citep{1995ApJ...442L..61S}. Ultraviolet (UV) and optical observations reveal an absence of significant hot wind plasma, and the winds are thus too cool to be Parker-type thermally driven flows \cite[e.g.,][]{1979ApJ...229L..27L,1980ApJ...235..519H,1981ApJ...250..293A}. 

Magnetic fields are most likely involved in the mass-loss process, although current magnetic models are also unable to explain spectral diagnostics. Exquisite high signal-to-noise ratio (S/N) \textit{Hubble} UV spectra have revealed that the one-dimensional (1 D) linear Alfv\'en-wave-driven wind models of the 1980’s (e.g., \citealt{1980ApJ...242..260H,1988PhDT........13H}) are untenable \citep{harper_2001}. These models predict chromospheres as integral parts of a turbulent, extended, and heated wind acceleration zone, but the theoretical line profiles and electron densities do not agree with the \textit{Hubble} spectra \cite[e.g.,][]{1998ApJ...494..828J}. One important property of cool evolved star winds gleaned from UV spectra is that, for the most part, the red giant winds accelerate in a quasi-steady manner and are not the result of ballistic ejecta as shown by the increase of wind scattering absorption velocity with optical depth in Fe\,\textsc{ii} lines \citep{1999ApJ...521..382C}. A new generation of theoretical models with outflows driven within diverging magnetic flux tubes has now emerged \citep{2006MNRAS.368.1145F, 2007ApJ...659.1592S}, but these too are not yet in agreement with observations \citep{2009AIPC.1094..267C}. It has also been suggested that the winds may be driven by some form of magnetic pressure acting on very highly clumped wind material \citep{2008AJ....136.1964E}, but \cite{2010ApJ...720.1767H} does not find compelling evidence for this hypothesis. Progress in this field continues to be driven by observations that provide new insights into the mass-loss problem.

\subsection{Radio Continuum Observations} \label{intro1} 

Although studies of wind-scattered UV and optical line profiles have provided clues to the mass-loss rates and radial distribution of the mean and turbulent velocity fields, the thermal structure remains poorly constrained. In the UV, the source function, $S_{\nu}$, of electron collisionally excited emission lines is sensitive to electron temperature, $T_{\rm{e}}$ (i.e., $S_{\nu} \propto e^{-h\nu /kT_{\rm{e}}}/\sqrt{T_{\rm{e}}}$). Therefore, a localized hot plasma component in a dynamic atmosphere can completely dominate the temporally and spatially averaged emission and hence not reflect the mean radial electron temperature distribution. At radio wavelengths, however, the source function is thermal and is just the Rayleigh-Jeans tail of the Planck function, which is linear in electron temperature (i.e., $S_{\nu} = {2kT_{\rm{e}}\nu ^2 /c^2}$). This should give a more appropriate estimate of the mean radial electron temperature. It is this value that controls the atomic level populations and ionization of the mean plasma, which is needed to quantify the implied thermal heating supplied to the  wind by the unknown driving source/sources, allowing constraints on potential mass-loss mechanisms to be derived.

In the centimeter-radio regime the radio opacity, $ \kappa _{\lambda}$, strongly increases with wavelength (i.e., $ \kappa _{\lambda} \propto \lambda ^{2.1}$), and so the longer wavelengths sample the extended layers of a star's atmosphere, thus	 providing us with spatial information about the star's mass outflow region. The NRAO\footnote{The National Radio Astronomy Observatory is a facility of the National Science Foundation operated under cooperative agreement by Associated Universities, Inc.} Karl G. Jansky Very Large Array (VLA) is sensitive to over three orders of magnitude in continuum optical depth, $\tau _{\lambda}$ ($\tau _{20 \ \rm{cm}}/\tau _{0.7 \ \rm{cm}} \approx 10^3$), and provides an area-averaged sweep through the wind acceleration zone of evolved late-type stars. The thermodynamic properties in this spatial region control the ionization in the far wind because the ionization balance, which also controls the cooling rates, becomes \textit{frozen-in} at large radii due to advection. Furthermore, it is these outer extended regions of the star's atmosphere that contribute to the commonly seen P Cygni line profiles in the UV. In these profiles the line-of-sight absorption caused by the star's wind is superimposed on the blueshifted scattered emission. Thus, centimeter radio continuum observations can provide a test of models based on these UV profiles. In this paper we directly compare our new VLA observations with atmospheric models derived from UV analysis.

\subsection{Sample Selection} \label{intro2}

\begin{deluxetable}{lcc}
\tabletypesize{\scriptsize}
\tablecaption{Stellar and Wind Parameters of $\alpha$ Boo and $\alpha$ Tau.}
\tablehead{	\colhead{}		       				& 
			\colhead{$\alpha$ Boo}				&
			\colhead{$\alpha$ Tau}				}
\startdata
Spectral type$\dotfill$	& K2 III  & K5 III  \\
HD number$\dotfill$				& 124897  & 29139  \\
Mass ($M_{\odot}$)$\dotfill$	& $0.8 \pm 0.2$  & $1.3 \pm 0.3$ \\
Effective temperature (K)$\dotfill$	& $4294 \pm 30$  & $3970 \pm 49$ \\
Angular diameter (mas)$\dotfill$		& $21.0 \pm 0.2$ & $20.2 \pm 0.3$ \\
Distance (pc)$\dotfill$	& $11.3 \pm 0.1$ & $20.4 \pm 0.3$\\
Radius ($R_{\odot}$)$\dotfill$	& $25.4 \pm 0.3$  & $44.4 \pm 1.0$ \\
Photospheric escape velocity$\dotfill$ & 110 km s$^{-1}$ & 106 km s$^{-1}$ \\
Rotation period (yr)$\dotfill$ & $2.0 \pm 0.2$ & 1.8 \\
$\lbrack$Fe/H$\rbrack\dotfill$ & $-0.5 \pm 0.2$ & $-0.15 \pm 0.2$  \\
Wind terminal velocity$\dotfill$ & $35 - 40$\,km\,s$^{-1}$ & 30\,km\,s$^{-1}$ \\
Mass-loss rate ($M_{\odot}$ yr$^{-1}$)$\dotfill$	& 2 $\times$ 10$^{-10}$ & 1.6 $\times$ 10$^{-11}$ \\
Wind temperature (K)$\dotfill$		& $\sim$10,000  & $\lesssim$10,000  \\
Semi-empirical model$\dotfill$	& \cite{1985pssl.proc..351D} & \cite{1999MNRAS.302...37M}
\enddata
\tablecomments{Masses are from \cite{2010A&A...509A..77K} and \cite{2012A&A...547A.108L}. Effective temperatures and photospheric angular diameters are from \cite{1993AA...270..315D}. Distances are from \cite{2007A&A...474..653V}. Rotaion periods are from \cite{2006PASP..118.1112G} and \cite{1993ApJ...413..339H}. Metallicities are from \cite{2003A&A...400..709D}. Wind parameters are derived from the semi-empirical models of \cite{1985pssl.proc..351D} and \cite{1998ApJ...503..396R}. The Ca\,\textsc{ii} ionization studies of \cite{2004IAUS..219..651H} indicate a wind temperature of $T_{e}\lesssim 1\times 10^4$ K for $\alpha$ Tau.}

\label{tab:tab1}
\end{deluxetable}

Currently the most detailed spatial information about the atmospheres of K and early M evolved stars is obtained from the $\zeta$ Aurigae and symbiotic eclipsing binaries (e.g., \citealt{1970VA.....12..147W}; \citealt{1996ApJ...466..979B}; \citealt{2008AJ....136.1964E}; \citealt{2008ApJ...675..711C}). Even though these systems offer us the best opportunity to obtain information on the dynamics and thermodynamics at various heights in the evolved star's atmosphere, the very nature of the binary system may introduce further complexities. For example, the orbital separation is often within the wind acceleration region, and one could expect flow perturbations to be present (e.g., \citealt{1981ApJ...248.1043C}). Using the ``old'' VLA, \cite{2005AJ....129.1018H} find a slow wind acceleration for $\zeta$ Aurigae and confirm that its velocity structure is not typical of single stars with similar spectral types, such as $\lambda$ Velorum \citep{1999ApJ...521..382C}. 

\begin{deluxetable*}{lccccccccc}
\tabletypesize{\scriptsize}
\tablecaption{VLA Observations}
\tablehead{\colhead{Star}			            &
\colhead{Date}
				&
          	\colhead{Band}	      				&
          	\colhead{Frequency\tablenotemark{a}}&
	\colhead{Wavelength}			         	&
           	\colhead{Time on Star}            &
           	\colhead{Restoring Beam}              &
           	\colhead{Bandwidth}            		&
           	\colhead{Number of}            		&
	\colhead{Phase}		\\
	\colhead{}									&
    \colhead{}		                			& 
    \colhead{}		                			& 
	\colhead{(GHz)}                        		& 
	\colhead{(cm)}                         		& 
	\colhead{(hr)}                    			&
	\colhead{($\arcsec \times \arcsec$ )}                    			&
	\colhead{(GHz)}                        		&
	\colhead{Antennae\tablenotemark{b}}         &
	\colhead{Calibrator}		}
\startdata
$\alpha$ Boo 	& 2011 Feb 22 & \textit{Q}	& 43.3 & 0.7		& 0.3 	&0.19 $\times$ 0.15& 0.256	&22& J1357+1919  \\
				& 2011 Feb 22 & Ka	& 33.6 & 0.9		& 0.2 	&0.25 $\times$ 0.20& 0.256 	&23&J1357+1919  \\
				& 2011 Feb 22 & \textit{K}	& 22.5 & 1.3		& 0.4	&0.35 $\times$ 0.28& 0.256 	&24&J1357+1919  \\
				& 2011 Feb 11 & \textit{X}	& 8.5  & 3.5		& 0.3 	&1.14 $\times$ 0.70& 0.256 	&18&J1415+1320  \\
				& 2011 Feb 11 & \textit{C}	& 5.0  & 6.0 		& 0.5	&2.02 $\times$ 1.30& 0.256 	&21& J1415+1320 \\
				& 2011 Feb 13 & \textit{S}	& 3.1  & 9.5 		& 1.8 	&2.57 $\times$ 2.08& 0.256 	&12& J1415+1320 \\
				& 2012 Jul 19 & \textit{S}	& 3.0  & 10.0 		& 0.7 	&2.82 $\times$ 2.30& 2.0		&23& J1415+1320 \\
				& 2012 Jul 20 & \textit{L}	& 1.5  & 20.0		& 1.6 	&4.46 $\times$ 3.94& 1.0		&23& J1415+1320 \\
\hline
\rule{0pt}{3ex}  $\alpha$ Tau	& 2011 Feb 11 & \textit{Q}	& 43.3 & 0.7 		& 0.3 	&0.18 $\times$ 0.16& 0.256 	&22&  J0431+1731\\
				& 2011 Feb 11 & Ka	& 33.6 & 0.9 		& 0.2 	&0.22 $\times$ 0.20& 0.256 	&19&  J0449+1121\\
				& 2011 Feb 11 & \textit{K}	& 22.5 & 1.3 		& 0.4 	&0.35 $\times$ 0.31& 0.256 	&21&  J0449+1121\\
				& 2011 Feb 13 & \textit{X}	&  8.5 & 3.5 		& 0.5	&0.85 $\times$ 0.78& 0.256 	&25&  J0449+1121\\
				& 2011 Feb 13 & \textit{C}	&  5.0 & 6.0 		& 1.2	&1.48 $\times$ 1.32& 0.256 	&21&  J0449+1121\\
				& 2011 Feb 12 & \textit{S}	&  3.1 & 9.5 		& 1.8 	&2.74 $\times$ 2.02& 0.256 	&11&  J0431+2037
\enddata
\tablenotetext{a}{Central frequency of selected bandpass.}
\tablenotetext{b}{Number of available antennae remaining after flagging.}
\label{tab:tab2}
\end{deluxetable*}

In order to avoid the assumed additional complexities of a companion, we have selected two single luminosity class III red giants: Arcturus ($\alpha$ Boo: K2 III) and Aldebaran ($\alpha$ Tau: K5 III). These nearby red giants have been extensively studied at other wavelengths, and their stellar parameters, which are briefly summarized in Table 1, are accurately known. Both of these late-type giants have ``hybrid atmospheres'' as they show evidence for both coronal/transition region activity and strong winds. Even though they are slow rotators, three possible values for the mean longitudinal magnetic field (albeit weak: $B = 0.65 \pm 0.26, 0.43 \pm 0.16$, and $-0.23 \pm 0.20$ G) have been reported for $\alpha$ Boo \citep{2011A&A...529A.100S} along with a possible magnetic cycle with a period of $\geq$14 yr \citep{2008ApJ...679.1531B}. Also, the detection of O $\textsc{vi}$ in $\alpha$ Tau \citep{2005ApJ...622..629D} indicates magnetic activity in its atmosphere. These stars are predicted to be point sources at all frequencies between 1 and 50 GHz in all VLA configurations, so our radio observations measure their total flux density, $F_{\nu}$. Moreover, both stars have existing semi-empirical 1 D chromospheric and wind models, which we directly compare to our data in this paper.

\section{OBSERVATIONS AND DATA REDUCTION}

Observations of $\alpha$ Boo and $\alpha$ Tau were carried out with the VLA during Open Shared Risk Observing in 2011 February at \textit{Q}, Ka, \textit{K}, \textit{X}, \textit{C}, and \textit{S} band in B-configuration (PI: G. M. Harper; Program ID: 10C-105). $\alpha$ Boo was also observed at \textit{S} and \textit{L} band in 2012 July when the VLA was again in B-configuration (PI: E. O'Gorman; Program ID: 12A-472). Some details of these observations are given in Table \ref{tab:tab2}. For the 2011 observations, the correlator was set up with two 128 MHz sub-bands centered on the frequencies listed in Table \ref{tab:tab2}. Each sub-band had 64 channels of width 2 MHz and four polarization products (RR, LL, RL, LR). For the \textit{S} and \textit{L} band observations in 2012, the $1 - 2$ GHz and $2 - 4$ GHz frequency ranges were both divided into 16 sub-bands, each with 64 channels. The channel width was 2 and 1 MHz for \textit{S} and \textit{L} band, respectively.

Both $\alpha$ Boo and $\alpha$ Tau were slightly offset from the phase center by $\sim$5 synthesized beam widths in order to avoid possible errors at phase-center. All scheduling blocks were kept to $\le$\,2.5 hr of duration. For the high-frequency observations (i.e., \textit{Q}, Ka, and \textit{K} bands) we used the \textit{fast switching} technique which consists of rapidly alternating observations of the target source and a nearby unresolved phase calibrator. The total cycle times for the \textit{Q}-, Ka-, and \textit{K}-band observations were 160, 230, and 290 s, respectively. For both target sources these high-frequency observations were combined into a single 2 hr observing track and commenced with \textit{X}-band reference pointing with solutions being applied on-line. After \textit{X}-band pointing the target source was observed at \textit{Q}-band to ensure that the best pointing solutions were used. The tracks at lower frequencies were composed of repeatedly interleaved observations of the target source and a nearby phase calibrator but had longer cycle times. The primary calibration sources 3C286 and 3C138 were observed at the end of all tracks and were used to measure the complex bandpass and set the absolute flux for $\alpha$ Boo and $\alpha$ Tau, respectively.  

\begin{deluxetable*}{ccccccccc}
\tabletypesize{\scriptsize}
\tablecaption{VLA Flux Densities of $\alpha$ Boo and $\alpha$ Tau}
\tablehead{\colhead{Star}		            &
			\colhead{Band}		            &
			\colhead{$\nu$\tablenotemark{a}}&
			\colhead{$\lambda$}&
			\colhead{Peak $F_{\nu}$}					&
          	\colhead{Integrated $F_{\nu}$} 			&
          	\colhead{\textit{Imfit} Integrated $F_{\nu}$}	&
			\colhead{Image rms}					&
           	\colhead{\textit{Imfit} Fitting Error}  	\\
	\colhead{}									&
	\colhead{}									&
	\colhead{(GHz)}								&
	\colhead{(cm)}								&
    \colhead{(mJy beam$^{-1}$)}		        	& 
    \colhead{(mJy)}		                		& 
	\colhead{(\rm{mJy})}                        		& 
	\colhead{(mJy beam$^{-1}$)}         					&
	\colhead{(mJy)}		}
\startdata
$\alpha$ Boo  &\textit{Q}  &43.28&0.7 &5.94 & 6.09 & 6.42 & 0.30 &  0.26\\
&Ka &33.56&0.9& 4.16 & 4.32 & 4.49 & 0.08 & 0.09 \\
&\textit{K}  &22.46&1.3& 1.83 & 1.78 & 1.81 & 0.04 & 0.05 \\
&\textit{X}  &8.46&3.5 & 0.51 & 0.51 & 0.53 & 0.03 & 0.02 \\
&\textit{C}  &4.90&6.1 & 0.21 & 0.14 & 0.16 & 0.04 & 0.01 \\
&\textit{S}  &3.15&9.5 & 0.15 & 0.14 & $\dots$     & 0.03 & $\dots$     \\
&\textit{S}  &2.87&10.4 & 0.13 & 0.12 & 0.12 & 0.01 & 0.02\\
&\textit{L}  &1.63&18.4 & 0.07 & 0.07 & $\dots$     & 0.01 & $\dots$    \\
\hline
\rule{0pt}{3ex}  $\alpha$ Tau &\textit{Q}  &43.28 & 0.7&3.67& 3.73 & 4.08 &  0.26& 0.18	\\
&Ka &33.56 & 0.9&2.19& 1.96 & 2.13 &  0.09& 0.07 \\
&\textit{K}  &22.46 &1.3& 1.86& 1.88 & 2.07 &  0.04& 0.08 \\
&\textit{X}  &8.46  & 3.5&0.30& 0.29 & 0.28 &  0.01& 0.02 \\
&\textit{C}  &4.96  & 6.0&0.15& 0.17 & 0.18 &  0.01& 0.01 \\
&\textit{S}  &3.15  & 9.5&0.06& 0.04 & $\dots$ &  0.02 & $\dots$
\enddata
\tablenotetext{a}{Frequency of the final image produced using the multi-frequency synthesis imaging mode within CASA's \textit{clean} task.}
\label{tab:tab3}
\end{deluxetable*}

The data were flagged, calibrated, and imaged within the Common Astronomical Software Application  \cite[CASA;][]{2007ASPC..376..127M} package. Data deemed to be bad by the VLA online system were flagged, as were zeros, non-operational antennae, dummy scans at the beginning of each track, and poorly performing antennae. Visual inspection of each scan was carried out to determine if data at the beginning or end of these scans needed to be flagged. For the 2011 low-frequency data the two sub-bands were centered at relatively radio frequency interference (RFI) free regions of the bandpass and only a very small amount of RFI had to be flagged.   The 2012 wide-band data were initially Hanning smoothed (combining adjacent frequency channels with weights 0.25, 0.5, and 0.25) to suppress Gibbs ringing. We manually flagged entire sub-bands that were badly contaminated with RFI. The \textit{testautoflag} task was then used to conservatively flag RFI from all sources and any remaining RFI was manually flagged. 

In order to calibrate the data, we solved for the complex gains of the calibration sources while applying the bandpass solution, which was derived from the relevant flux calibrator.  The amplitude gains of the phase calibrators were scaled according to values derived from the flux calibrators using the ``Perley-Butler 2010" flux density standard \citep{2013ApJS..204...19P}. At the time, no Ka- or \textit{S}-band flux density standard models were available, so instead for these we used the \textit{K} and \textit{L}-band models, respectively, which were scaled according to their spectral indices. The more frequently observed phase calibrators were then used to calibrate the amplitude and phases of the targets. Atmospheric opacity corrections were also applied to the high frequency data sets using the average of a seasonal model (based on many years of measurements) and information from the weather station obtained during the observations.

The visibilities were then both Fourier transformed and deconvolved using the CASA \textit{clean} task in multi-frequency synthesis imaging mode, which separately grids the multiple spectral channels onto the \textit{u-v} plane and therefore improves the overall \textit{u-v} coverage. We used natural weighting for maximum sensitivity, and the cell size was chosen so that the synthesized beam was about five pixels across. For the high frequencies it was usually sufficient to place just one CLEAN circle around the target source.  For the low frequencies, however, the image sizes were usually set to a few times the size of the primary beam so that nearby strong serendipitous sources could be CLEANed, thus reducing their sidelobe contamination of the final image. These images were CLEANed interactively, taking sky curvature into account, down to about the $3\sigma$  level with clean boxes placed around sources as they appeared in the residual image. All images were corrected for  primary beam attenuation. 

In each image the flux density from the unresolved target source was calculated by, (1) taking the peak pixel value from the source, (2) manually integrating the flux density around the source, and (3) fitting an elliptical Gaussian model to the source and deriving the integrated flux density using the CASA \textit{imfit} task. Each of these values along with the image rms noise measured from adjacent background regions and fitting error produced by \textit{imfit} are given in Table \ref{tab:tab3} to indicate the quality of each radio map. Both sources are point sources at all frequencies, so the peak flux value given in Table \ref{tab:tab3} will also be its total flux density value. For weak detections (i.e., $F_{\nu} \lesssim 5\sigma$) we avoid using the \textit{imfit} task to obtain a flux density estimate, as this may produce biased parameter estimates \citep{1999ASPC..180.....T}. The flux density values used in Section \ref{disc:disc0} are the peak values listed in Table \ref{tab:tab3}. We assume absolute flux density scale systematic uncertainties of  3\% at all frequencies \citep{2013ApJS..204...19P}.

\section{RESULTS} 

Apart from $\alpha$ Boo at \textit{C} band and $\alpha$ Tau at \textit{S} band, detections were made in every sub-band for the 2011 data. For all other bands, the flux densities of the targets in both sub-bands were found to be the same within their uncertainties, so we do not present separate values here. Instead we give the values from the radio maps produced by concatenating the two sub-bands. We present in Table \ref{tab:tab3} the target flux densities extracted from these concatenated radio maps. In the following two sections we briefly discuss the properties of these radio maps for both targets.

\subsection{$\alpha$ Boo Radio Maps} \label{results1} 
High-S/N detections ($>$19$\sigma$) of $\alpha$ Boo were made at 22.5, 33.6, and 43.3 GHz. Some residuals of the dirty beam remained in the CLEANed maps due to the paucity of uv-coverage in these short high-frequency observations. At the lower frequencies, it was necessary to image confusing sources, notably a strong radio source located $186\arcsec$ north-west of $\alpha$ Boo. This non-thermal source was reported by \cite{1986AJ.....91..602D}, and their flux density of 25 mJy at 4.9 GHz is in close agreement with our measurement of 23.2 mJy at the same frequency. We find the source to have a spectral index $\alpha$ ($F_{\nu} \propto \nu ^{\alpha}$) of $-1.4$ between 8.5 and 1.6 GHz; its flux density reaches 80.3 mJy at 1.6 GHz.

\begin{deluxetable*}{ccccccc}
\tabletypesize{\scriptsize}
\tablecaption{Compilation of Previous Radio Observations of $\alpha$ Boo and $\alpha$ Tau ($\nu \le 250$ GHz)}
\tablehead{							            &
    \colhead{$\nu$ (GHz)}		    			& 
		    \colhead{$\lambda$ (cm)}			&
			\colhead{Date}						&
			\colhead{$F_{\nu}$ (mJy)}			&
			\colhead{S/N}				&
          	\colhead{Source} 					}
\startdata
$\alpha$ Boo (K2 III) &4.9  & 6.1&1983 Jan 21 & 0.39 & 3.0 & \cite{1986AJ.....91..602D} \\
&4.9  &6.1& 1983 May 20 & 0.26 & 3.3& \cite{1986AJ.....91..602D} \\
&4.9  &6.1& 1983 Dec 26 & $\le$0.18$(3\sigma)$& $\dots$ & \cite{1986AJ.....91..602D} \\
&4.9  &6.1& 1984 Mar 17 & 0.24  & 4.8& \cite{1986AJ.....91..602D} \\
&15.0 &2.0& 1984 Nov 6 & 0.68 & 7.6& \cite{1986AJ.....91..602D} \\
&22.5  &1.3& 1999 Jan 6  &1.7& 8.5& \cite{2011AA...533A.107D} \\
&43.3  &0.69& 1999 Jan 6 & 3.3& 8.3& \cite{2011AA...533A.107D} \\
&43.3  &0.69& 2004 Jan 25 & 3.34& 41.8& \cite{2011AA...533A.107D} \\
&86.0  &0.35& 1985 Nov  & 21.4& 3.0& \cite{1986AA...164..227A} \\
&108.4  &0.28& 1997 Nov - 2000 Jun & 20.1 &29.1 & \cite{2005AJ....129.2836C} \\
&217.8 &0.14& 1997 Nov - 2000 Jun  & 83.5 &48.8 & \cite{2005AJ....129.2836C} \\
&250.0  &0.12& 1986 Dec - 1989 Mar  & 78.0 & 9.8& \cite{1994AA...281..161A} \\
\hline
\rule{0pt}{3ex}    $\alpha$ Tau (K5 III)	&4.9&6.1& 1983 Jan 21 & $\le$0.27$(3\sigma)$&$\dots$& \cite{1986AJ.....91..602D} \\
&4.9&6.1  & 1984 Nov 6 & $\le$0.22$(3\sigma)$&$\dots$& \cite{1986AJ.....91..602D} \\
&5.0  & 6.0&1997 Sep 27 & $\le$0.07$(3\sigma)$	&$\dots$& \cite{2007ApJ...655..946W} \\
&8.5  &3.5& 1997 Sep 27 & 0.28 	&9.3	& \cite{2007ApJ...655..946W} \\
&14.9 &2.0& 1997 Sep 27 & 0.95 	&11.9	& \cite{2007ApJ...655..946W} \\
&15.0 &2.0& 1984 Nov 6 & 0.60 	&6.0	& \cite{1986AJ.....91..602D} \\
&108.4 &0.28 & 1997 Nov - 2000 Dec &  14.0  & 9.6& \cite{2005AJ....129.2836C} \\
&217.8 &0.14& 1999 Sep - 2000 Dec  & 25.8 & 4.6& \cite{2005AJ....129.2836C} \\
&250.0 &0.12 & 1986 Dec - 1987 Jan & 51.0 & 8.5& \cite{1994AA...281..161A} 
\enddata
\label{tab:tab4}
\end{deluxetable*}

We detected $\alpha$ Boo at 6$\sigma$ in the lower frequency sub-band of \textit{C} band, at 4.9 GHz. The noise was slightly higher and the images were poorer quality in the \textit{C}-band higher frequency sub-band, with artifacts exceeding $\pm$200 $\mu$Jy, and we cannot report a detection in this sub-band, so values given in Table \ref{tab:tab3} are taken from the lower frequency sub-band only. We obtain good detections ($>$5$\sigma$) of the star for both epochs at $\sim$3 GHz (\textit{S} band), and the peak flux densities agree within their uncertainties. We can therefore safely assume that the 1.5 GHz (\textit{L} band) flux density has not changed significantly over that period either, and so can safely be included in any analysis. The map at \textit{L} band was highly contaminated by the sidelobes of the strong source north-west of $\alpha$ Boo but the star is still detected at the $5\sigma$ level. There is a slight positional offset of $1\arcsec$ between the position of the peak flux density at 1.5 and at 3.0 GHz for the 2012 data, which were taken within 1 day of each other. However, the position uncertainties due to noise and phase uncertainties between the directions of the phase reference source and the target are at least $1\arcsec$, and so we feel that it is highly likely that both detections are of $\alpha$ Boo. 

\subsection{$\alpha$ Tau Radio Maps} \label{results2}
The final deconvolved radio maps of $\alpha$ Tau were of excellent quality with the rms noise reaching the predicted noise levels in many cases. The target field at all frequencies was free from strong serendipitous radio sources, and thus the final images were free of the sidelobe contamination that was present in the low-frequency $\alpha$ Boo images. $\alpha$ Tau was the only source in the high-frequency maps, while the brightest source in the low-frequency maps was located $106\arcsec$ north-north east of $\alpha$ Tau and had flux densities of 0.85, 1.35, and 1.7 mJy at 8.5, 5.0, and 3.5 GHz, respectively. Strong detections ($>$14$\sigma$) of $\alpha$ Tau were made at all frequencies between 5.0 and 43.3 GHz. Due to the limited number of \textit{S}-band receivers available at the time, a full 2.5 hr track was dedicated to $\alpha$ Tau at 3.1 GHz in order to achieve the required sensitivity to give a possible detection. We report a tentative $3\sigma$ detection of $\alpha$ Tau at 3.1 GHz when we take its peak pixel value as its total flux density. 

\section{DISCUSSION} \label{disc:disc0}
\subsection{Results versus Previous Observations} \label{disc1}
\begin{figure*}
\centering
\includegraphics[trim = 0mm 0mm 0mm 0mm, clip,scale=0.62, angle=90]{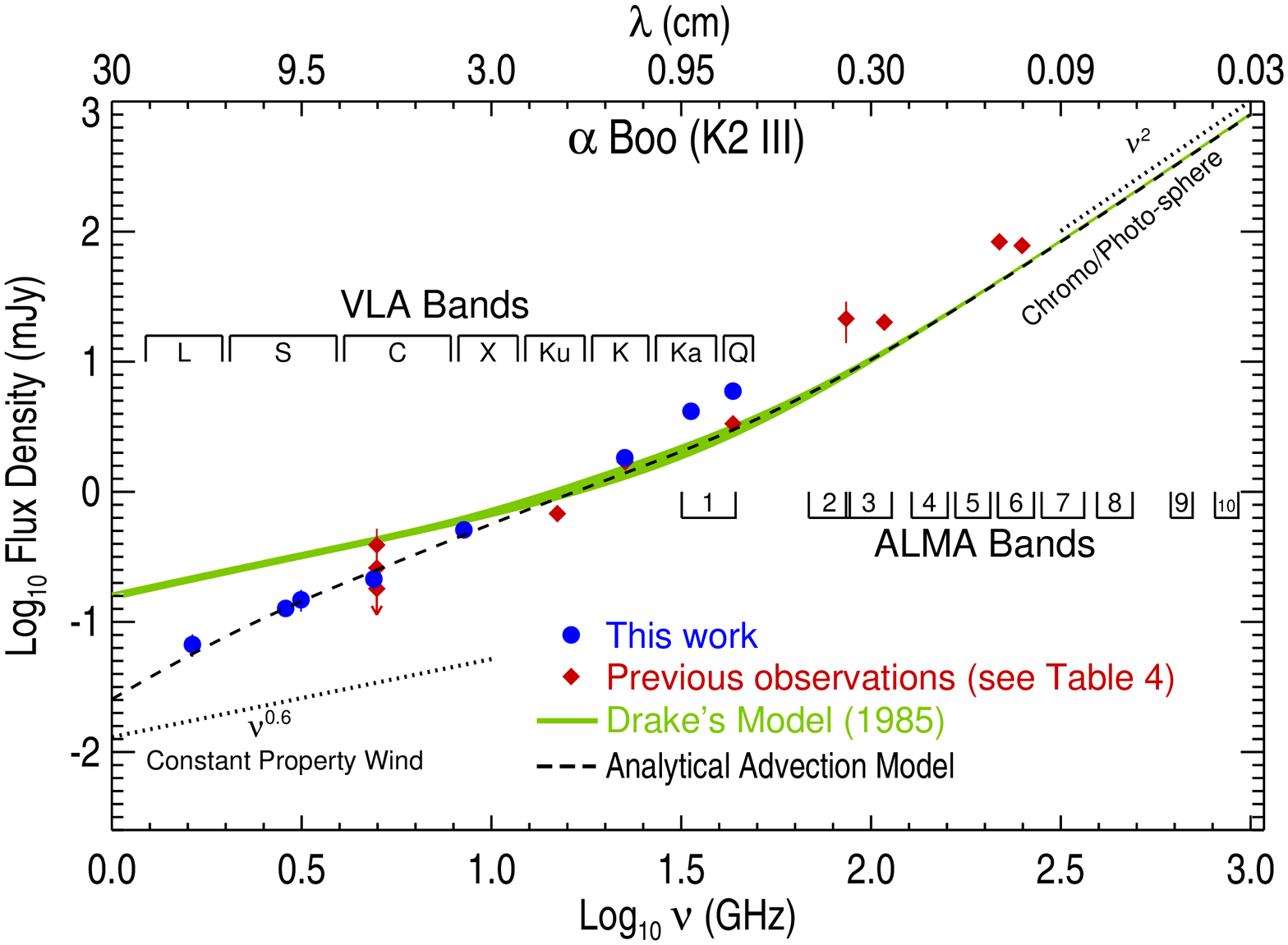}
\\
\caption{Spectral energy distribution of $\alpha$ Boo for 1 GHz $\leq \nu \leq$ 1 THz. Our new multi-frequency VLA observations which were mainly acquired over a few days in 2011 February are the blue circles and disagree with the existing chromospheric and wind models of \cite{1985pssl.proc..351D}. The overlap between the two models is represented by the green shaded area. The red diamonds are previous observations which were acquired sporadically over the past three decades with the ``old'' VLA, IRAM, and BIMA. The black dashed line is the expected radio emission from the Drake model which undergoes rapid wind cooling beyond $\sim$2.3 $R_{\star}$ (see Sections \ref{disc:disc3} and \ref{disc:disc4}).}
\label{fig:fig1}
\centering
\includegraphics[trim = 0mm 0mm 0mm 0mm, clip,scale=0.62, angle=90]{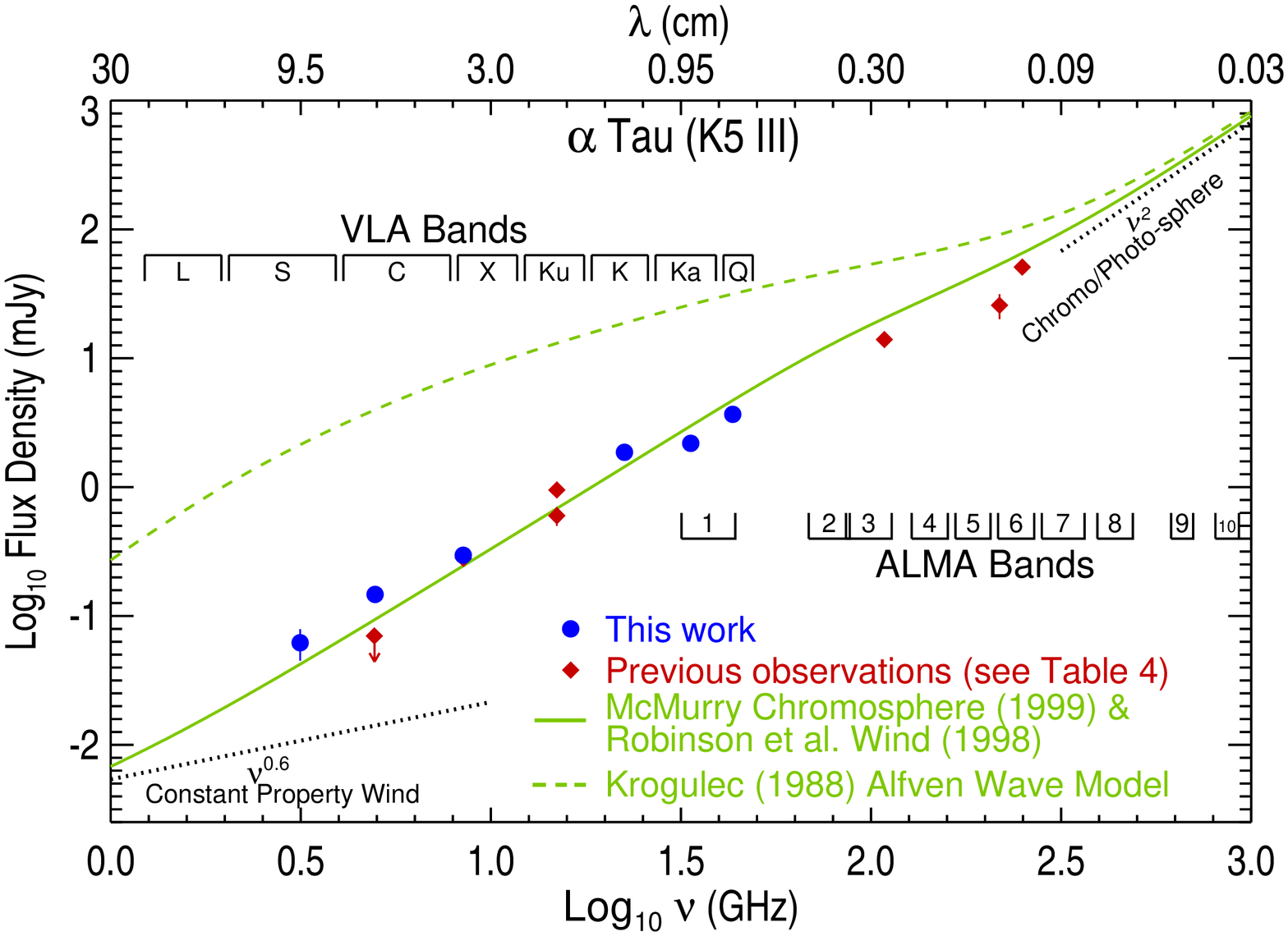}
\\
\caption{Spectral energy distribution of $\alpha$ Tau for 1 GHz $\leq \nu \leq$ 1 THz. Our new multi-frequency VLA observations of $\alpha$ Tau (blue circles) were acquired in just two days in 2011 February. The red diamonds are the previous radio observations of the star which were acquired over many years (see Table \ref{tab:tab4}). The green line is the expected radio emission from the existing hybrid chromosphere and wind model, while the dashed  green line is the expected radio emission from a theoretical Alfv\'en-wave-driven model atmosphere.}
\label{fig:fig2}
\end{figure*}

Prior to and during the early operation of the ``old'' VLA, a small number of single dish radio observations reported the detection of flares from single red giants (e.g., \citealt{1989MNRAS.239..913S}). These transient radio events have never been re-observed, however, even with more sensitive interferometers, suggesting that such detections were spurious (e.g., \citealt{1992MNRAS.254....1B}). The first definitive detection of thermal free-free emission from a luminosity class III single red giant at centimeter wavelengths was of $\alpha$ Boo at 6 cm \citep{1983ApJ...274L..77D,1986AJ.....91..602D}. Since then there has been a modest number of centimeter and millimeter observations of this star. In Table \ref{tab:tab4} we list the majority of these observations and plot their flux densities as a function of frequency in Figure \ref{fig:fig1}. In comparison to other single red giants, $\alpha$ Boo had been relatively well observed at radio continuum wavelengths before this study, including detections in four VLA bands (i.e., \textit{Q}, \textit{K}, Ku, and \textit{C}). No Ku-band receivers were available during the commissioning phase of the VLA in early 2011, so we can compare three of our detections with previous ones. 

Previous detections of $\alpha$ Boo at 6 cm ranged from a 3$\sigma$ upper limit of 0.18 mJy to a 3$\sigma$ detection at 0.39 mJy. Our 6 cm value agrees to within $\sim$10$\%$ of the highest S/N (5$\sigma$) value of \cite{1986AJ.....91..602D}. There is no significant difference between our 1.3 cm value and that of \cite{2011AA...533A.107D}. There is however a notable difference in flux density values at 0.7 cm  where \cite{2011AA...533A.107D} report values that are lower than ours by over 40\%. Although we do not rule out such a level of chromospheric radio variability, it is not expected based on the small level of UV variability observed from such supposedly inactive stars \citep{2013MNRAS.428.2064H}. Another possibility for the difference in values is that the longer cycle time used by \cite{2011AA...533A.107D}, which was over double our value, may lead to larger phase errors and thus lower final flux density values. Future high-frequency VLA observations of $\alpha$ Boo will clarify this discrepancy at 0.7 cm, but past detections at longer wavelengths appear to be in good agreement with our data.

In Figure \ref{fig:fig2} we plot the previous  radio measurements of $\alpha$ Tau at all frequencies below 250 GHz (i.e., $> 0.12$ cm). Prior to this study, $\alpha$ Tau had only been detected at two VLA bands (i.e., X and Ku) and had never been detected at wavelengths longer than 3 cm due to its relatively low mass-loss rate. Our lack of a Ku-band measurement means that we can only compare the previous 3 cm detection reported in \cite{2007ApJ...655..946W} to ours. We find that there is no significant difference between the two. Interestingly, \cite{2007ApJ...655..946W} report a non-detection of $\alpha$ Tau at 6 cm and placed a 3$\sigma$ upper limit of 0.07 mJy on its emission. In stark contrast to this, we were able to detect the star at 6 cm with a flux density over two times greater than this value. This hint of variability at long wavelengths would be consistent with the predictions of the broadband nonlinear Alfv\'{e}n wave model of \cite{2010ApJ...723.1210A} but can only be confirmed with future high-S/N observations.

\subsection{Existing Atmospheric Models} \label{disc2}
\begin{figure}
\includegraphics[trim = 0mm 0mm 0mm 0mm, clip,scale=0.36,angle=90]{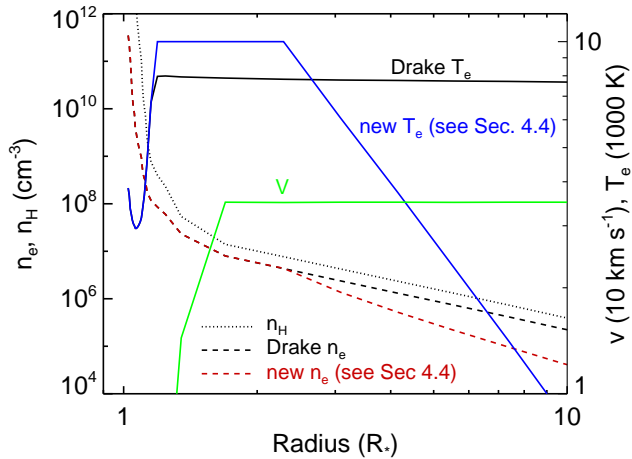}
\caption{Existing atmospheric model for $\alpha$ Boo \cite[``model A'']{1985pssl.proc..351D} along with the same model which undergoes rapid wind cooling beyond $\sim$2.3 $R_{\star}$ (see Section \ref{disc:disc4}). The original Drake models have a temperature plateau of $\sim$8,000 K between 1.2 and $\sim$20 $R_{\star}$ (solid black line), reach a terminal velocity of $35 - 40$ km s$^{-1}$ within 2 $R_{\star}$ (solid green line), and have a wind which is 50\% ionized (dashed and dotted black lines).}
\label{fig:fig0}
\end{figure}

One of the most important diagnostic features indicating mass outflows in late-type evolved stars are the blue shifted absorption components present in the Ca\,\textsc{ii} H and K and Mg\,\textsc{ii} h and k resonance lines. Figure \ref{fig:fig0} shows one of the two chromosphere and wind models of $\alpha$ Boo \cite[``model A'']{1985pssl.proc..351D} which is based on the Mg\,\textsc{ii} k $\lambda$2796 emission line observed with the \textit{IUE} telescope. The line was modeled by solving the radiative transfer equation in a spherical co-moving frame and the effects of partial redistribution \citep[e.g.,][]{1983ApJ...273..299D} were taken into account. Both of Drake's atmospheric models are semi-empirical and contain no assumptions about the wind-driving mechanism. They contain the photospheric model of \cite{1975ApJ...200..660A}, predict the wind to reach a terminal velocity of $35 - 40$ km s${}^{-1}$ by 2 $R _{\star}$, and reach a maximum microturbulence of 5 km s$^{-1}$. They contain a broad temperature plateau with $T_{\rm{e}}$ $\approx$ 8000 K between 1.2 and $\sim$20 $R _{\star}$ with a cooler region farther out, and hydrogen is $\sim50$\% ionized. We compute the radio spectrum from these models assuming spherical 1-D geometry \citep{1994MNRAS.268..894H} with the free-free Gaunt factors from \cite{1988ApJ...327..477H}. The radiative transfer equation is solved using the Feautrier technique \citep{1978stat.book.....M}, and the boundary condition is determined by ensuring that the atmosphere is optically thick at the deepest layers. \cite{1985pssl.proc..351D} predicts that their atmospheric model would produce a flux density value of 0.4 mJy at 6 cm, and encouragingly, our radio spectrum reproduces this value. Departures from spherical symmetry are to be expected in magnetic stellar atmospheres. For example, $\alpha$ Boo has an inclination axis of 58$^{\circ} \pm$25$^{\circ}$ \citep{2006PASP..118.1112G}, and a global magnetic dipole could cause density variations between the equator and the polar regions. Despite this fact, the study of a  spherically symmetric atmosphere forms the basis of understanding the more complex environments in real stellar atmospheres.

Figure \ref{fig:fig1} shows the resulting predicted radio spectrum between 1 GHz and 1 THz for $\alpha$ Boo from these chromosphere and wind models (green line). At high frequencies the radio spectra produced by these models have a blackbody-like slope (i.e., $\sim \nu ^{2}$) as a result of the small ion density scale heights close to the star where the temperature is changing slowly. At low frequencies however, where the Drake models predict the wind to have constant velocity, ionization fraction, and temperature, the slope approaches the well known $\sim\nu ^{0.6}$ limit \citep{1975MNRAS.170...41W,1975AA....39..217O,1975AA....39....1P}. The paucity and, in some cases, low S/N of previous observations made it difficult to discern the validity of this model prior to our multi-frequency study of $\alpha$ Boo. Our new data reveal significant deviations from the semi-empirical model at both low and high frequencies (in this case below $\sim$8 GHz and above $\sim$25 GHz). At high frequencies our VLA data indicate a flux excess which is in agreement with previous mm observations. This may be due to larger chromospheric ion densities or to the possible presence of transition region plasma not accounted for in the Drake model. The discrepancy at low frequencies may be due to a lower ionization fraction in the wind, or a lower mass-loss rate than that used in the Drake model.

In Figure \ref{fig:fig2} we plot the expected radio spectrum of $\alpha$ Tau based on the semi-empirical 1-D chromosphere and transition region model of \cite{1999MNRAS.302...37M} embedded in the 1-D wind model of \cite{1998ApJ...503..396R}. The semi-empirical McMurry model was created by using the radiative transfer code MULTI \citep{carlsson_1986} to reproduce the fluxes of collisionally excited C\,\textsc{i}, C\,\textsc{ii}, Si\,\textsc{iii}, Mg\,\textsc{ii}, and C\,\textsc{iv} lines in a plane-parallel, hydrostatic, one-component atmosphere. It contains the photospheric model of \cite{1973ApJ...180...81J} and reaches a maximum temperature of 10$^{5}$ K at 1.2 $R_{\star}$. As it does not contain a wind outflow, we use Robinson et al.'s wind characteristics beyond 1.2 $R_{\star}$ to describe the outflow velocity. In this wind model, the wind reaches $\sim$80\% of its terminal value of 30 km s$^{-1}$ by 3 $R_{\star}$. The Robinson et al. wind characteristics are based on matching the Fe\,\textsc{ii} 2755 $\rm{\AA}$ line and the O\,\textsc{i} triplet near $\lambda$1304 with a simplified wind model using the SEI computer code \citep{1987ApJ...314..726L}. We assume the wind to have a constant temperature of 10,000 K and have a constant ionization fraction of 0.6 throughout, based on the ionization fraction at the corresponding temperature in the McMurry model.

The radio flux densities at high frequencies (i.e., $\nu >$ 30 GHz) are overestimated by the combination of both atmospheric models, although this approach does well in reproducing the VLA flux densities below 30 GHz. The VLA, Institut de Radioastronomie Millim\'{e}trique (IRAM) 30 m telescope, and Berkeley Illinois Maryland Association (BIMA) continuum flux densities confirm that this model predicts a flux excess at even higher frequencies. One possible explanation for this is that the inner atmosphere contains extensive amounts of cooler gas than that predicted by the 1-D static chromosphere and transition region model of McMurry. This scenario agrees with the findings of \cite{1994ApJ...423..806W}, who conclude that cool regions exist close to the stellar surface with large ($> 99 \%$) filling factors, i.e., a thermally bifurcated CO-mosphere \citep{1996IAUS..176..371A}. The wind which we have overlain on top of the McMurry chromosphere and transition region is found to be optically thin at nearly all VLA wavelengths, and only contributes a very small flux at the longest wavelengths. As our model matches the data reasonably well below 30 GHz, we conclude that $\alpha$ Tau's wind is optically thin and the VLA radio emission at all wavelengths emanates from the inner atmosphere. 

We also include the predicted radio spectrum from the theoretical Alfv\'{e}n-wave-driven outflow model for $\alpha$ Tau \citep{1989AcA....39..51K} in Figure \ref{fig:fig2} to demonstrate how radio observations can empirically challenge theoretical models. This model has a fully ionized outflow inside 10 $R_{\star}$ and has a mass-loss rate of 6.3 $\times$ 10$^{-9}$ $M_{\odot}$ yr$^{-1}$, more than two orders of magnitude higher than the more recent estimate given in Table \ref{tab:tab1}. As the radio opacity is proportional to $n _{\rm{e}} n _{\rm{ion}}$, where $n_{\rm{e}}$ and $n_{\rm{ion}}$ are the electron and ion number densities, respectively, this model greatly overestimates the actual radio flux density at all VLA wavelengths. The linear Alfv\'{e}n wave models for $\alpha$ Boo \citep{1988AcA....38..107K} also assume full ionization and have higher mass-loss rates than the value given in Table \ref{tab:tab1}, predicting higher flux densities than observed. The lack of agreement between the Alfv\'en-wave-driven wind models of \cite{1988AcA....38..107K,1989AcA....39..51K} and our observed radio fluxes may not necessarily be due to an incorrect  wind-driving mechanism and instead may be due to the simplifications and  uncertainties in these models, such as wind densities, magnetic field strengths, damping lengths, and flow geometries close to the star. For example, the mass-loss rate is very sensitive to the radial surface magnetic field strength (i.e., $\dot{M} \propto B^4$) in these Alfv\'en wave models \citep{1983ApJ...275..808H} so a small uncertainty in the mass-loss rate can lead to a large uncertainty in the magnetic field strength. Relaxing some of these simplifications such as purely radial flows or non-assumption of the WKB (Wentzel-Kramers-Brillouin) approximation \citep{1995ApJ...454..901C} may also lead to better agreement with our radio data.

Recently, \cite{2013A&A...553A...3O} has detected a layer of CO in the outer atmosphere of $\alpha$ Tau (i.e., a so-called MOLsphere) which extends out to $\sim$2.5 $R_{\star}$, has a temperature of $1500 \ \pm 200$ K, and has a CO column density of $\sim$1$\times 10^{20}$ cm$^{-2}$. They were unable to constrain the geometrical thickness, $\Delta L$, of the MOLsphere from the data, however, and arbitrarily set it to 0.1 $R_{\star}$. It can be shown that such a MOLsphere would have an optical depth of $\tau _{\rm{6\,cm}} = 4.6$ at \textit{C} band and would produce a corresponding flux density of  0.06 mJy, which is considerably lower than our high-S/N measurement of 0.15 mJy. Here, we have conservatively assumed that the electrons in this region of the atmosphere come from singly ionized metals and have an abundance of $\sim$1$\times 10^{-5} n_{\rm{H}}$, where $n_{\rm{H}}$ is the total hydrogen number density. We have also assumed the CO filling factor to be unity. The disagreement in values between our radio data and the predicted flux for an optically thick disk could mean that the MOLsphere is optically thin at long VLA wavelengths and that the radio emission emanates from the more ionized material closer to the star. It can be shown that the MOLsphere becomes optically thin at \textit{C} band (i.e., $\tau _{\rm{6\,cm}} < 1$) for $\Delta L >$0.46 $R_{\star}$, so if the long VLA wavelength radio emission from $\alpha$ Tau comes from a region closer in to the star, then the MOLsphere either has a geometrical width  $>$0.46 $R _{\star}$ or has a filling factor less than unity.

\subsection{Radio Spectral Indices} \label{disc:disc3}
Long-wavelength radio emission from non-dusty K spectral-type red giants is due to thermal free-free emission in their partially ionized outflows, while shorter wavelength radio emission emanates from nearly static lower atmospheric layers. The radio flux density-frequency relationship for these stars is usually found to be intermediate between that expected from the isothermal stellar disk emission, where $\alpha$ follows the Rayleigh-Jeans tail of the Planck function (i.e., $\alpha = +2$), and that from an optically thin plasma ($\alpha = -0.1$). It can be shown that the expected radio spectrum from a spherically symmetric isothermal outflow with a constant velocity and ionization fraction varies as $\nu ^{0.6}$ \citep{1975MNRAS.170...41W,1975AA....39..217O,1975AA....39....1P}. If we relax some of the assumptions about the outflow in this constant property wind model and instead assume that the electron density and temperature vary as a function of distance from the star $r$, and have the power-law form $n_{e}(r) \propto r^{-p}$ and $T_{e}(r) \propto r^{-n}$, respectively, then 

\begin{figure}
\includegraphics[trim = 0mm 0mm 0mm 10mm, clip,scale=0.385,angle=90]{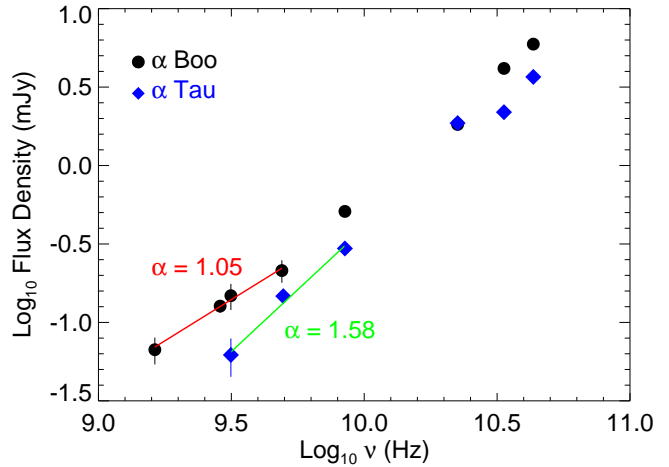}
\caption{Radio spectra for $\alpha$ Boo and $\alpha$ Tau, together with the best fit power law to their long-wavelength flux densities and the resulting spectral indices. The spectral indices for $\alpha$ Boo and $\alpha$ Tau are found to be 1.05 and 1.58, respectively, which are both larger than the 0.6 value expected for a constant property wind model.}
\label{fig:fig3}
\end{figure}

\begin{equation}
\alpha = \frac{4p -6.2 -0.6n}{2p-1-1.35n}
\label{eq:eq1}
\end{equation}
\citep[e.g.,][]{1987ApJ...312..813S}. These power law approximations are only going to be valid over certain radial ranges of the star's outflow.

The radio spectra for both stars are shown in Figure \ref{fig:fig3}, together with the power-laws that were fitted to the long wavelength flux densities by minimizing the chi-square error statistic. For $\alpha$ Boo a power law with $F_{\nu} \propto \nu ^{1.05 \pm 0.05}$ fits the four longest wavelength data points well. This spectral index is larger than the 0.8 value obtained by \cite{1986AJ.....91..602D} whose value was based on a shorter wavelength (2 cm) value and a mean value of four low S/N measurements at 6 cm. $\alpha$ Tau was found to have a larger spectral index and a power law with $F_{\nu} \propto \nu ^{1.58 \pm 0.25}$ best fitted the three longest wavelength data points. This value is in agreement with \cite{1986AJ.....91..602D}, who report a value $\ge 0.84$, and is lower than the value of 2.18 that can be derived from the shorter wavelength data given in \cite{2007ApJ...655..946W}. It should be emphasized that the spectral index for both stars is steeper than that expected from the constant property wind model. 

\begin{figure}
\includegraphics[trim = 0mm 11mm 10mm 21mm, clip,scale=0.4,angle=90]{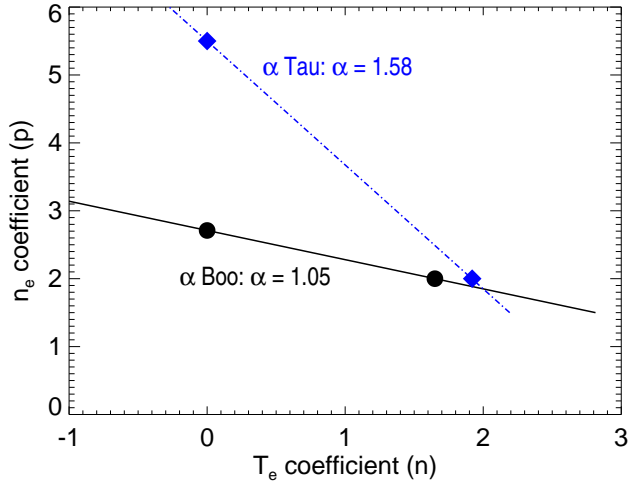}
\caption{Variation of density and temperature coefficients for the empirically derived spectral indices. The density coefficients for an isothermal flow ($n=0$) along with the temperature coefficients for a constant outflow velocity ($p=2$) are also shown for both stars.}
\label{fig:fig4}
\end{figure}

Equation (1) can be used in conjunction with our new spectral index for each star to calculate the density and temperature coefficients that may describe their outflows. The combinations of the electron temperature and density coefficients are shown for each star in Figure 5 ($\alpha$ Boo is represented by the solid line and $\alpha$ Tau by the dash-dotted line) along with the coefficients obtained by assuming either an isothermal flow ($n=0$) or a constant velocity flow ($p=2$). One potential explanation for spectral indices of stellar outflows being larger than 0.6 is that the wind is still accelerating in the radio emitting region, if the thermal gradients are assumed to be small. For an isothermal flow, the density coefficients are $p  =2.7$ and 5.5 for $\alpha$ Boo and $\alpha$ Tau, respectively. From mass conservation, and assuming a steady flow, the power-law coefficients for the velocity profiles of each star can be found, i.e., $v(r) \propto r^{p-2}$. For $\alpha$ Boo we find $v(r) \propto r^{0.7}$, while for $\alpha$ Tau we find $v(r) \propto r^{3.5}$. This suggests that our long VLA wavelengths may probe a steep acceleration region for $\alpha$ Tau's outflow but for $\alpha$ Boo, may probe a region where the wind is close to its terminal velocity. 

\begin{figure}
\includegraphics[trim = 5mm 10mm 10mm 20mm, clip,scale=0.4,angle=90]{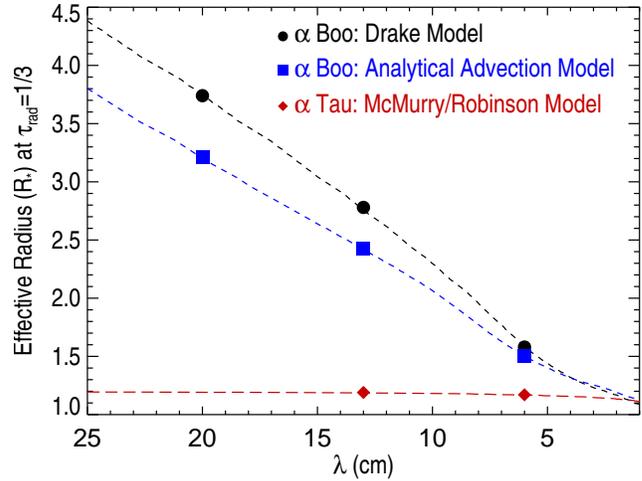}
\caption{Predicted effective radius (dashed lines) as a function of wavelength derived from the existing atmospheric models of $\alpha$ Boo and $\alpha$ Tau.  Also plotted is the predicted effective radius derived from our analytical advection model for $\alpha$ Boo (discussed in Section \ref{disc:disc4}). Points corresponding to our long-wavelength VLA measurements are also shown. At the same radio wavelengths the lower mass-loss rate of $\alpha$ Tau results in a smaller effective radius than that for $\alpha$ Boo.}
\label{fig:fig5}
\end{figure}

The assumption of shallow thermal gradients in a stellar outflow is probably unreliable, however. It is likely that some form of Alfv\`{e}n waves are required to lift the material out of the gravitational potential as suggested by \cite{1980ApJ...242..260H}. These waves need damping lengths which are much larger than the chromospheric density scale height $H$ (where $H \sim 0.01\ R_{\star}$ for our targets), in order to lift the material out of the gravitational potential, but $\lesssim 1 \, R_{\star}$ in order to avoid wind terminal velocities greater than those observed \citep[e.g.,][]{1983ApJ...275..808H}. It has also been shown that these waves are expected to produce substantial heating near the base of the wind \citep[e.g.,][]{1982ApJ...261..279H}. If the long-wavelength radio emission from $\alpha$ Tau is indeed emanating from the wind acceleration region, then the dissipation of these Alfv\`{e}n waves may introduce thermal gradients in this region and its velocity profile will not be described by $v(r) \propto r^{3.5}$. Furthermore, diverging flow geometries have been invoked as a more realistic representation of stellar atmospheres \citep{1982ApJ...257..264H,1989A&A...209..327J,2006SSRv..122..181V}, and so one could write the area of a flux tube (normalized to its value at $r_{1}$) as $(r/r_{1})^2f(r)$, where $f(r)$ is a function describing the divergence from a purely radial flow [i.e., when $f(r)=1$].  If the non-radial expansion term can be described by a power law $f(r) \propto r^s$, where $s>1$ indicates super-radial expansion, then $v(r) \propto r^{3.5-s}$ would describe the velocity profile of $\alpha$ Tau (assuming an isothermal flow with a constant ionization fraction). Therefore, including diverging geometries reduces the magnitude of the acceleration. Farther out in the wind, where it has reached its terminal velocity, one would also expect a thermal gradient (but now of opposite sign) due to adiabatic expansion and line cooling. If the long-wavelength radio emission emanates from this region of the wind, then Equation (1) provides us with a direct estimate of the temperature coefficient as we can assume the density coefficient is $p=2$. This may be the case for our long VLA wavelength measurements of $\alpha$ Boo, in which case $T_{e}(r) \propto r^{-1.65}$.

To investigate this matter further,  we estimate the effective radius of the radio emitting region as a function of wavelength based on the Drake model for $\alpha$ Boo and the hybrid McMurry and Robinson et al. model for $\alpha$ Tau. We follow the approach used by \cite{1977ApJ...212..488C} and assume that the radio emission at each wavelength is characterized by emission from a radial optical depth $\tau _{\rm{rad}}\sim 1/3$. This is a modification of the Eddington-Barbier relation for an extended atmosphere where emission from smaller optical depths has added weight. Since the radio free-free opacity increases at longer wavelengths, the optical depth along a line of sight into the stellar outflow also increases at longer wavelengths. This implies that the effective radius (i.e., the radius where $\tau _{\lambda} = \tau _{\rm{rad}}$) will increase with longer wavelengths and will be greater for outflows with higher densities of ionized material as $\tau _{\lambda}(r) \propto \lambda ^{2.1} \int n_{\rm{ion}}(r)n_{\rm{e}}(r) dr$. 

The larger mass-loss rate of $\alpha$ Boo in comparison to $\alpha$ Tau means that the latter has a substantially smaller effective radius at longer wavelengths, as seen in Figure \ref{fig:fig5}. At 6, 13, and 20 cm the effective radius of $\alpha$ Boo at $\tau _{\rm{rad}}$=1/3 is predicted to be 1.6, 2.8, and 3.7 $R_{\star}$ but is only $\sim$1.2 $R_{\star}$ at 6 and 13 cm for $\alpha$ Tau. \cite{1998ApJ...503..396R} predict that $\alpha$ Tau's wind reaches $\sim$80\% of its terminal velocity by 3 $R_{\star}$, but even our longest-wavelength observations are highly unlikely to sample the wind outside the lower velocity layers closer to the star. For $\alpha$ Boo, however, \cite{1985pssl.proc..351D} predicts that the wind has reached its terminal velocity by $\sim$2 $R_{\star}$, so based on this model our longest-wavelength measurements are of the region where the wind has reached a steady terminal velocity. From Figure \ref{fig:fig4}, this implies that the $n_{\rm{e}}$ coefficient is $p=2$ and thus the $T_{\rm{e}}$ coefficient is $n=1.65$. Pure adiabatic spherical expansion cooling with no heat source has $n=1.33$, so additional cooling routes must be operating, possibly due to line cooling. Finally, the wind ionization balance may not have become \textit{frozen-in} in the region of $\alpha$ Boo's wind where the radio emission emanates from. If this is true, then the excess slope of the spectral index could be due to a combination of both cooling and changing ionization fraction. In this scenario the temperature coefficient $n$, would be smaller than our derived value because Equation (1) assumes a constant ionization fraction.

\subsection{Analytical Advection Model for $\alpha$ Boo's Wind} \label{disc:disc4}

A failure of the Drake model for $\alpha$ Boo is that it overestimates the radio fluxes at long VLA wavelengths which sample the outer atmosphere, as clearly shown in Figure \ref{fig:fig1}. If these wavelengths are indeed sampling the wind at its terminal velocity, then one reason for this overestimation is that the wind is cooling closer in than predicted by the existing model, which assumes a constant temperature of 8000 K out to $\sim$20 $R_{\star}$. The main mechanism for such cooling would be adiabatic expansion \citep{2011ASPC..448..691O} and would cause lower electron densities than those predicted by the existing model due to larger recombination rates. 

To investigate the possibility of the wind undergoing more rapid cooling closer in to the star, we adjusted one of the existing models (referred to as ``Model A'' in \citealt{1985pssl.proc..351D}) to include a temperature power-law falloff of the form
\begin{equation}
T_{e}(r)= T_{e}(r_{1})\left(\frac{r_{1}}{r}\right)^{n},
\label{eq:eq2}
\end{equation}
at some distance $r_{1}$ from the star, and used the temperature coefficient $n=1.65$ obtained from our new VLA data assuming a constant velocity flow (see Figure \ref{fig:fig4}). We introduce the distance $r_{1}$ as the outer limit to ionization processes; at $r > r_{1}$, the ionization fraction is only determined by recombination. To calculate the new electron density in the wind regime where this temperature falloff occurs, we used the analytical expression of \cite{1986ApJ...306..605G} to calculate the hydrogen ionization fraction, $x_{\rm{H\,\textsc{ii}}}=n_{\rm{H\,\textsc{ii}}}/n_{\rm{H}}$, where $n_{\rm{HII}}$ and $n_{\rm{H}}$ are the ionized and total hydrogen number densities, respectively. To do so, we need to make a number of assumptions about the wind properties beyond radius $r_{1}$, namely:

1. A constant velocity mass outflow, i.e., $n_{\rm{H}}(r)=C/r^2$, where $C$ is a constant proportional to the ratio of the mass-loss rate divided by the terminal velocity. For $\alpha$ Boo, $C = 1.7 \times 10^{32} $ cm$^{-1}$ assuming a wind velocity of 35 km s$^{-1}$.

2. All hydrogen ionization processes cease beyond $r_{1}$. The ionization of hydrogen in the chromosphere and wind is a two-stage process: the $n = 2$ level is excited by electron collisions and Ly$\alpha$ scattering, followed by photoionization by the optically thin Balmer continuum. When the temperature begins to decrease in the wind the collisional excitation rate and thus ionization rate decrease rapidly.

3. Only radiative recombination of hydrogen is considered and the temperature variation of the recombination coefficient $\alpha _{B}$ which excludes captures to the $n=1$ level \citep{1978ppim.book.....S} is included. The recombination coefficient varies with temperature as
\begin{equation}
\alpha _{B}(r) = \alpha _{B}(r_{1})\left[\frac{T_{e}(r_1)}{T_{e}(r)}\right]^{0.77},
\label{eq:eq3}
\end{equation}
where the power-law coefficient is obtained by finding the slope of the recombination coefficients between 1000 K and 16,000 K given in \cite{1978ppim.book.....S}.

4. A fixed ion contribution from metals with a low first ionization potential, $x_{\rm{ion}}=n_{\rm{ion}}/n_{\rm{H}}=10^{-4}$, as these are easily ionized in the outflow.\\
Using these assumptions, it can be shown that the ionization fraction beyond $r_{1}$ is given by \citep{1986ApJ...306..605G}
\begin{equation}
x_{\rm{H\,\textsc{ii}}}(r)= \frac{x_{\rm{H\,\textsc{ii}}}(r_1)x_{\rm{ion}}e^{-I(r)}}{x_{\rm{ion}}+x_{\rm{H\,\textsc{ii}}}(r_1)[1 - e^{-I(r)}]}
\label{eq:eq4}
\end{equation}
where
\begin{equation}
I(r) = \frac{4.7\times 10^9}{r_{1}}\left[\left( \frac{r_{1}}{r}\right)^{-0.27} -1 \right], \ \rm{and} \ \  r \geq r_{1}.
\label{eq:eq5}
\end{equation}

We adjusted the value of $r_{1}$ to obtain the best fit to our long wavelength observations and found that this happened when $r_{1}$ = 2.3 $R_{\star}$. To get this best fit, the existing atmospheric model (plotted in Figure \ref{fig:fig0}) needed to be adjusted, so that it now has a narrower and slightly larger temperature plateau of $T_e = 10,000$ K between 1.2 and 2.3 $R_{\star}$ and a temperature profile and a density profile governed by Equations (2) and (4) beyond $r_{1}$ = 2.3 $R_{\star}$, respectively. This gives good agreement with our new long wavelength VLA data as shown in Figure \ref{fig:fig1}. This new \textit{hybrid} model, which is plotted along with the original Drake model in Figure \ref{fig:fig0}, still has the original ionization fraction of $x_{\rm{H\,\textsc{ii}}} \approx 0.5$ inside 2.3 $R_{\star}$ but now contains an initial rapid decrease in $x_{\rm{H\,\textsc{ii}}}$ beyond 2.3 $R_{\star}$, which then \textit{freezes-in} to a constant value of $\sim$0.04 beyond $\sim$10 $R_{\star}$.

Encouraging as it is that such a simple analytical model can reproduce values close to the observed radio fluxes at long wavelengths, it must be stressed that this \textit{hybrid} model is just a first-order approximation. It assumes that the excess slope from the radio spectrum is a result of rapid cooling only. It still does not reproduce the radio fluxes at wavelengths shorter than $\sim$3 cm, and therefore a new atmospheric model is still required that can reproduce all of the observed flux densities. To do so, the non-trivial task of simultaneously solving the radiative transfer equation and non-LTE atomic level populations which include advection will be required.

\section{CONCLUSIONS}
We have presented the most comprehensive set of multi-wavelength radio continuum observations of two standard luminosity class III red giants to date. This is the first time such stars have been detected at wavelengths longer than 6 cm. Such long-wavelength detections are crucial if one wants to study the outer environments of these partially ionized stellar outflows. Our observations were carried out with the VLA during its commissioning phase when only a fraction of the now available bandwidth was at our disposal. The continuous bandwidth coverage between 1 and 50 GHz of the new VLA will allow fast detections of historically weak or undetectable radio continuum luminosity class III red giants at both long and short wavelengths. Previous upper limits will be replaced by firm detections, allowing a greater understanding of their outer atmospheric properties.

The spectral index of both $\alpha$ Tau and $\alpha$ Boo at long wavelengths is found to be greater than that expected from a constant property wind. For $\alpha$ Tau our longest wavelength detections are still sampling emission from an accelerating region within the outflow, while for $\alpha$ Boo the emission probably emanates from a region where the flow is close to, or indeed has reached, its terminal velocity. Using our new VLA data, we have developed a simple analytical model for the outer atmosphere of $\alpha$ Boo which contains a rapid wind cooling profile. Future detailed non-LTE radiative transfer models which include advection are required to match all radio flux densities at all wavelengths. 

\acknowledgments
The data presented in this paper were obtained with the Karl G. Jansky Very Large Array (VLA), which is an instrument of the National Radio Astronomy Observatory (NRAO). The NRAO is a facility of the National Science Foundation operated under cooperative agreement by Associated Universities, Inc. We wish to thank the NRAO helpdesk for their detailed responses to our CASA-related queries. We thank the referee for their careful reading of the manuscript and their valuable comments. This publication has emanated from research conducted with the financial support of Science Foundation Ireland under Grant Number SFI11/RFP.1/AST/3064, and a grant from Trinity College Dublin.

{\it Facility:} \facility{VLA}.

\bibliography{references}

\begin{thebibliography}{79}
\expandafter\ifx\csname natexlab\endcsname\relax\def\natexlab#1{#1}\fi

\bibitem[{{Airapetian} {et~al.}(2010){Airapetian}, {Carpenter}, \&
  {Ofman}}]{2010ApJ...723.1210A}
{Airapetian}, V., {Carpenter}, K.~G., \& {Ofman}, L. 2010, \apj, 723, 1210

\bibitem[{{Altenhoff} {et~al.}(1986){Altenhoff}, {Huchtmeier}, {Schmidt},
  {Schraml}, \& {Stumpff}}]{1986AA...164..227A}
{Altenhoff}, W.~J., {Huchtmeier}, W.~K., {Schmidt}, J., {Schraml}, J.~B., \&
  {Stumpff}, P. 1986, \aap, 164, 227

\bibitem[{{Altenhoff} {et~al.}(1994){Altenhoff}, {Thum}, \&
  {Wendker}}]{1994AA...281..161A}
{Altenhoff}, W.~J., {Thum}, C., \& {Wendker}, H.~J. 1994, \aap, 281, 161

\bibitem[{{Ayres}(1996)}]{1996IAUS..176..371A}
{Ayres}, T.~R. 1996, in IAU Symposium, Vol. 176, Stellar Surface Structure, ed.
  K.~G. {Strassmeier} \& J.~L. {Linsky}, 371

\bibitem[{{Ayres} {et~al.}(1997){Ayres}, {Brown}, {Harper}, {Bennett},
  {Linsky}, {Carpenter}, \& {Robinson}}]{1997ApJ...491..876A}
{Ayres}, T.~R., {Brown}, A., {Harper}, G.~M., {Bennett}, P.~D., {Linsky},
  J.~L., {Carpenter}, K.~G., \& {Robinson}, R.~D. 1997, \apj, 491, 876

\bibitem[{{Ayres} \& {Linsky}(1975)}]{1975ApJ...200..660A}
{Ayres}, T.~R., \& {Linsky}, J.~L. 1975, \apj, 200, 660

\bibitem[{{Ayres} {et~al.}(1981){Ayres}, {Linsky}, {Vaiana}, {Golub}, \&
  {Rosner}}]{1981ApJ...250..293A}
{Ayres}, T.~R., {Linsky}, J.~L., {Vaiana}, G.~S., {Golub}, L., \& {Rosner}, R.
  1981, \apj, 250, 293

\bibitem[{{Baade} {et~al.}(1996){Baade}, {Kirsch}, {Reimers}, {Toussaint},
  {Bennett}, {Brown}, \& {Harper}}]{1996ApJ...466..979B}
{Baade}, R., {Kirsch}, T., {Reimers}, D., {Toussaint}, F., {Bennett}, P.~D.,
  {Brown}, A., \& {Harper}, G.~M. 1996, \apj, 466, 979

\bibitem[{{Beasley} {et~al.}(1992){Beasley}, {Stewart}, \&
  {Carter}}]{1992MNRAS.254....1B}
{Beasley}, A.~J., {Stewart}, R.~T., \& {Carter}, B.~D. 1992, \mnras, 254, 1

\bibitem[{{Brown} {et~al.}(2008){Brown}, {Gray}, \&
  {Baliunas}}]{2008ApJ...679.1531B}
{Brown}, K.~I.~T., {Gray}, D.~F., \& {Baliunas}, S.~L. 2008, \apj, 679, 1531

\bibitem[{{Carlsson}(1986)}]{carlsson_1986}
{Carlsson}, M. 1986, Uppsala Astronomical Observatory Reports, 33

\bibitem[{{Carpenter} {et~al.}(1999){Carpenter}, {Robinson}, {Harper},
  {Bennett}, {Brown}, \& {Mullan}}]{1999ApJ...521..382C}
{Carpenter}, K.~G., {Robinson}, R.~D., {Harper}, G.~M., {Bennett}, P.~D.,
  {Brown}, A., \& {Mullan}, D.~J. 1999, \apj, 521, 382

\bibitem[{{Cassinelli} \& {Hartmann}(1977)}]{1977ApJ...212..488C}
{Cassinelli}, J.~P., \& {Hartmann}, L. 1977, \apj, 212, 488

\bibitem[{{Chapman}(1981)}]{1981ApJ...248.1043C}
{Chapman}, R.~D. 1981, \apj, 248, 1043

\bibitem[{{Charbonneau} \& {MacGregor}(1995)}]{1995ApJ...454..901C}
{Charbonneau}, P., \& {MacGregor}, K.~B. 1995, \apj, 454, 901

\bibitem[{{Cohen} {et~al.}(2005){Cohen}, {Carbon}, {Welch}, {Lim}, {Schulz},
  {McMurry}, {Forster}, \& {Goorvitch}}]{2005AJ....129.2836C}
{Cohen}, M., {Carbon}, D.~F., {Welch}, W.~J., {Lim}, T., {Schulz}, B.,
  {McMurry}, A.~D., {Forster}, J.~R., \& {Goorvitch}, D. 2005, \aj, 129, 2836

\bibitem[{{Crowley} {et~al.}(2009){Crowley}, {Espey}, {Harper}, \&
  {Roche}}]{2009AIPC.1094..267C}
{Crowley}, C., {Espey}, B.~R., {Harper}, G.~M., \& {Roche}, J. 2009, in
  American Institute of Physics Conference Series, Vol. 1094, 15th Cambridge
  Workshop on Cool Stars, Stellar Systems, and the Sun, ed. E.~{Stempels},
  267--274

\bibitem[{{Crowley} {et~al.}(2008){Crowley}, {Espey}, \&
  {McCandliss}}]{2008ApJ...675..711C}
{Crowley}, C., {Espey}, B.~R., \& {McCandliss}, S.~R. 2008, \apj, 675, 711

\bibitem[{{Decin} {et~al.}(2003){Decin}, {Vandenbussche}, {Waelkens}, {Decin},
  {Eriksson}, {Gustafsson}, {Plez}, \& {Sauval}}]{2003A&A...400..709D}
{Decin}, L., {Vandenbussche}, B., {Waelkens}, C., {Decin}, G., {Eriksson}, K.,
  {Gustafsson}, B., {Plez}, B., \& {Sauval}, A.~J. 2003, \aap, 400, 709

\bibitem[{{Dehaes} {et~al.}(2011){Dehaes}, {Bauwens}, {Decin}, {Eriksson},
  {Raskin}, {Butler}, {Dowell}, {Ali}, \& {Blommaert}}]{2011AA...533A.107D}
{Dehaes}, S., {et~al.} 2011, \aap, 533, A107

\bibitem[{{di Benedetto}(1993)}]{1993AA...270..315D}
{di Benedetto}, G.~P. 1993, \aap, 270, 315

\bibitem[{{Drake} \& {Linsky}(1986)}]{1986AJ.....91..602D}
{Drake}, S., \& {Linsky}, J. 1986, \aj, 91, 602

\bibitem[{{Drake}(1985)}]{1985pssl.proc..351D}
{Drake}, S.~A. 1985, in Progress in stellar spectral line formation theory;
  Proceedings of the Advanced Research Workshop, Trieste, Italy, September 4-7,
  1984 (A86-37976 17-90). Dordrecht, D. Reidel Publishing Co., 1985, p.
  351-357., ed. J.~E. {Beckman} \& L.~{Crivellari}, 351--357

\bibitem[{{Drake} \& {Linsky}(1983{\natexlab{a}})}]{1983ApJ...274L..77D}
{Drake}, S.~A., \& {Linsky}, J.~L. 1983{\natexlab{a}}, \apjl, 274, L77

\bibitem[{{Drake} \& {Linsky}(1983{\natexlab{b}})}]{1983ApJ...273..299D}
---. 1983{\natexlab{b}}, \apj, 273, 299

\bibitem[{{Dupree} {et~al.}(2005){Dupree}, {Lobel}, {Young}, {Ake}, {Linsky},
  \& {Redfield}}]{2005ApJ...622..629D}
{Dupree}, A.~K., {Lobel}, A., {Young}, P.~R., {Ake}, T.~B., {Linsky}, J.~L., \&
  {Redfield}, S. 2005, \apj, 622, 629

\bibitem[{{Eaton}(2008)}]{2008AJ....136.1964E}
{Eaton}, J.~A. 2008, \aj, 136, 1964

\bibitem[{{Falceta-Gon{\c c}alves} {et~al.}(2006){Falceta-Gon{\c c}alves},
  {Vidotto}, \& {Jatenco-Pereira}}]{2006MNRAS.368.1145F}
{Falceta-Gon{\c c}alves}, D., {Vidotto}, A.~A., \& {Jatenco-Pereira}, V. 2006,
  \mnras, 368, 1145

\bibitem[{{Glassgold} \& {Huggins}(1986)}]{1986ApJ...306..605G}
{Glassgold}, A.~E., \& {Huggins}, P.~J. 1986, \apj, 306, 605

\bibitem[{{Gray} \& {Brown}(2006)}]{2006PASP..118.1112G}
{Gray}, D.~F., \& {Brown}, K.~I.~T. 2006, \pasp, 118, 1112

\bibitem[{{Haisch} {et~al.}(1980){Haisch}, {Linsky}, \&
  {Basri}}]{1980ApJ...235..519H}
{Haisch}, B.~M., {Linsky}, J.~L., \& {Basri}, G.~S. 1980, \apj, 235, 519

\bibitem[{{Harper}(1988)}]{1988PhDT........13H}
{Harper}, G.~M. 1988, PhD thesis, Oxford Univ.~(England).

\bibitem[{{Harper}(1994)}]{1994MNRAS.268..894H}
---. 1994, \mnras, 268, 894

\bibitem[{{Harper}(2001)}]{harper_2001}
{Harper}, G.~M. 2001, in Astronomical Society of the Pacific Conference Series,
  Vol. 223, 11th Cambridge Workshop on Cool Stars, Stellar Systems and the Sun,
  ed. R.~J. {Garcia Lopez}, R.~{Rebolo}, \& M.~R. {Zapaterio Osorio}, 368

\bibitem[{{Harper}(2010)}]{2010ApJ...720.1767H}
---. 2010, \apj, 720, 1767

\bibitem[{{Harper} {et~al.}(2004){Harper}, {Brown}, {Ayres}, \&
  {Sim}}]{2004IAUS..219..651H}
{Harper}, G.~M., {Brown}, A., {Ayres}, T., \& {Sim}, S.~A. 2004, in IAU
  Symposium, Vol. 219, Stars as Suns : Activity, Evolution and Planets, ed.
  A.~K. {Dupree} \& A.~O. {Benz}, 651

\bibitem[{{Harper} {et~al.}(2005){Harper}, {Brown}, {Bennett}, {Baade},
  {Walder}, \& {Hummel}}]{2005AJ....129.1018H}
{Harper}, G.~M., {Brown}, A., {Bennett}, P.~D., {Baade}, R., {Walder}, R., \&
  {Hummel}, C.~A. 2005, \aj, 129, 1018

\bibitem[{{Harper} {et~al.}(2013){Harper}, {O'Riain}, \&
  {Ayres}}]{2013MNRAS.428.2064H}
{Harper}, G.~M., {O'Riain}, N., \& {Ayres}, T.~R. 2013, \mnras, 428, 2064

\bibitem[{{Hartmann} {et~al.}(1982){Hartmann}, {Avrett}, \&
  {Edwards}}]{1982ApJ...261..279H}
{Hartmann}, L., {Avrett}, E., \& {Edwards}, S. 1982, \apj, 261, 279

\bibitem[{{Hartmann} \& {MacGregor}(1980)}]{1980ApJ...242..260H}
{Hartmann}, L., \& {MacGregor}, K.~B. 1980, \apj, 242, 260

\bibitem[{{Hartmann} \& {MacGregor}(1982)}]{1982ApJ...257..264H}
---. 1982, \apj, 257, 264

\bibitem[{{Hatzes} \& {Cochran}(1993)}]{1993ApJ...413..339H}
{Hatzes}, A.~P., \& {Cochran}, W.~D. 1993, \apj, 413, 339

\bibitem[{{Holzer} {et~al.}(1983){Holzer}, {Fla}, \&
  {Leer}}]{1983ApJ...275..808H}
{Holzer}, T.~E., {Fla}, T., \& {Leer}, E. 1983, \apj, 275, 808

\bibitem[{{Holzer} \& {MacGregor}(1985)}]{1985ASSL..117..229H}
{Holzer}, T.~E., \& {MacGregor}, K.~B. 1985, in Astrophysics and Space Science
  Library, Vol. 117, Mass Loss from Red Giants, ed. M.~{Morris} \&
  B.~{Zuckerman}, 229--255

\bibitem[{{Hummer}(1988)}]{1988ApJ...327..477H}
{Hummer}, D.~G. 1988, \apj, 327, 477

\bibitem[{{Jatenco-Pereira} \& {Opher}(1989)}]{1989A&A...209..327J}
{Jatenco-Pereira}, V., \& {Opher}, R. 1989, \aap, 209, 327

\bibitem[{{Johnson}(1973)}]{1973ApJ...180...81J}
{Johnson}, H.~R. 1973, \apj, 180, 81

\bibitem[{{Jones}(2008)}]{2008MNRAS.387..845J}
{Jones}, M.~H. 2008, \mnras, 387, 845

\bibitem[{{Judge} \& {Carpenter}(1998)}]{1998ApJ...494..828J}
{Judge}, P.~G., \& {Carpenter}, K.~G. 1998, \apj, 494, 828

\bibitem[{{Kallinger} {et~al.}(2010){Kallinger}, {Weiss}, {Barban}, {Baudin},
  {Cameron}, {Carrier}, {De Ridder}, {Goupil}, {Gruberbauer}, {Hatzes},
  {Hekker}, {Samadi}, \& {Deleuil}}]{2010A&A...509A..77K}
{Kallinger}, T., {et~al.} 2010, \aap, 509, A77

\bibitem[{{Krogulec}(1988)}]{1988AcA....38..107K}
{Krogulec}, M. 1988, Acta Astronomica, 38, 107

\bibitem[{{Krogulec}(1989)}]{1989AcA....39..51K}
---. 1989, Acta Astronomica, 39, 51

\bibitem[{{Lamers} {et~al.}(1987){Lamers}, {Cerruti-Sola}, \&
  {Perinotto}}]{1987ApJ...314..726L}
{Lamers}, H.~J.~G.~L.~M., {Cerruti-Sola}, M., \& {Perinotto}, M. 1987, \apj,
  314, 726

\bibitem[{{Lebzelter} {et~al.}(2012){Lebzelter}, {Heiter}, {Abia}, {Eriksson},
  {Ireland}, {Neilson}, {Nowotny}, {Maldonado}, {Merle}, {Peterson}, {Plez},
  {Short}, {Wahlgren}, {Worley}, {Aringer}, {Bladh}, {de Laverny}, {Goswami},
  {Mora}, {Norris}, {Recio-Blanco}, {Scholz}, {Th{\'e}venin}, {Tsuji},
  {Kordopatis}, {Montesinos}, \& {Wing}}]{2012A&A...547A.108L}
{Lebzelter}, T., {et~al.} 2012, \aap, 547, A108

\bibitem[{{Linsky} \& {Haisch}(1979)}]{1979ApJ...229L..27L}
{Linsky}, J.~L., \& {Haisch}, B.~M. 1979, \apjl, 229, L27

\bibitem[{{McMullin} {et~al.}(2007){McMullin}, {Waters}, {Schiebel}, {Young},
  \& {Golap}}]{2007ASPC..376..127M}
{McMullin}, J.~P., {Waters}, B., {Schiebel}, D., {Young}, W., \& {Golap}, K.
  2007, in Astronomical Society of the Pacific Conference Series, Vol. 376,
  Astronomical Data Analysis Software and Systems XVI, ed. R.~A. {Shaw},
  F.~{Hill}, \& D.~J. {Bell}, 127

\bibitem[{{McMurry}(1999)}]{1999MNRAS.302...37M}
{McMurry}, A.~D. 1999, \mnras, 302, 37

\bibitem[{{Mihalas}(1978)}]{1978stat.book.....M}
{Mihalas}, D. 1978, Stellar atmospheres 2nd edition (San Francisco,
  W.~H.~Freeman and Co., 1978.~650 p.)

\bibitem[{{O'Gorman} \& {Harper}(2011)}]{2011ASPC..448..691O}
{O'Gorman}, E., \& {Harper}, G.~M. 2011, in Astronomical Society of the Pacific
  Conference Series, Vol. 448, 16th Cambridge Workshop on Cool Stars, Stellar
  Systems, and the Sun, ed. C.~{Johns-Krull}, M.~K. {Browning}, \& A.~A.
  {West}, 691

\bibitem[{{Ohnaka}(2013)}]{2013A&A...553A...3O}
{Ohnaka}, K. 2013, \aap, 553, A3

\bibitem[{{Olnon}(1975)}]{1975AA....39..217O}
{Olnon}, F.~M. 1975, \aap, 39, 217

\bibitem[{{Panagia} \& {Felli}(1975)}]{1975AA....39....1P}
{Panagia}, N., \& {Felli}, M. 1975, \aap, 39, 1

\bibitem[{{Perley} \& {Butler}(2013)}]{2013ApJS..204...19P}
{Perley}, R.~A., \& {Butler}, B.~J. 2013, \apjs, 204, 19

\bibitem[{{Reimers}(1982)}]{1982A&A...107..292R}
{Reimers}, D. 1982, \aap, 107, 292

\bibitem[{{Robinson} {et~al.}(1998){Robinson}, {Carpenter}, \&
  {Brown}}]{1998ApJ...503..396R}
{Robinson}, R.~D., {Carpenter}, K.~G., \& {Brown}, A. 1998, \apj, 503, 396

\bibitem[{{Seaquist} \& {Taylor}(1987)}]{1987ApJ...312..813S}
{Seaquist}, E.~R., \& {Taylor}, A.~R. 1987, \apj, 312, 813

\bibitem[{{Sennhauser} \& {Berdyugina}(2011)}]{2011A&A...529A.100S}
{Sennhauser}, C., \& {Berdyugina}, S.~V. 2011, \aap, 529, A100

\bibitem[{{Slee} {et~al.}(1989){Slee}, {Stewart}, {Bunton}, {Beasley},
  {Carter}, \& {Nelson}}]{1989MNRAS.239..913S}
{Slee}, O.~B., {Stewart}, R.~T., {Bunton}, J.~D., {Beasley}, A.~J., {Carter},
  B.~D., \& {Nelson}, G.~J. 1989, \mnras, 239, 913

\bibitem[{{Spitzer}(1978)}]{1978ppim.book.....S}
{Spitzer}, L. 1978, Physical processes in the interstellar medium (New York
  Wiley-Interscience, 1978.\ 333 p.)

\bibitem[{{Sutmann} \& {Cuntz}(1995)}]{1995ApJ...442L..61S}
{Sutmann}, G., \& {Cuntz}, M. 1995, \apjl, 442, L61

\bibitem[{{Suzuki}(2007)}]{2007ApJ...659.1592S}
{Suzuki}, T.~K. 2007, \apj, 659, 1592

\bibitem[{{Taylor} {et~al.}(1999){Taylor}, {Carilli}, \&
  {Perley}}]{1999ASPC..180.....T}
{Taylor}, G.~B., {Carilli}, C.~L., \& {Perley}, R.~A., eds. 1999, Astronomical
  Society of the Pacific Conference Series, Vol. 180, {Synthesis Imaging in
  Radio Astronomy II}

\bibitem[{{van Leeuwen}(2007)}]{2007A&A...474..653V}
{van Leeuwen}, F. 2007, \aap, 474, 653

\bibitem[{{Vidotto} {et~al.}(2006){Vidotto}, {Falceta-Gon{\c c}alves}, \&
  {Jatenco-Pereira}}]{2006SSRv..122..181V}
{Vidotto}, A.~A., {Falceta-Gon{\c c}alves}, D., \& {Jatenco-Pereira}, V. 2006,
  \ssr, 122, 181

\bibitem[{{Wiedemann} {et~al.}(1994){Wiedemann}, {Ayres}, {Jennings}, \&
  {Saar}}]{1994ApJ...423..806W}
{Wiedemann}, G., {Ayres}, T.~R., {Jennings}, D.~E., \& {Saar}, S.~H. 1994,
  \apj, 423, 806

\bibitem[{{Wood} {et~al.}(2007){Wood}, {Harper}, {M{\"u}ller}, {Heerikhuisen},
  \& {Zank}}]{2007ApJ...655..946W}
{Wood}, B.~E., {Harper}, G.~M., {M{\"u}ller}, H.-R., {Heerikhuisen}, J., \&
  {Zank}, G.~P. 2007, \apj, 655, 946

\bibitem[{{Wright} \& {Barlow}(1975)}]{1975MNRAS.170...41W}
{Wright}, A.~E., \& {Barlow}, M.~J. 1975, \mnras, 170, 41

\bibitem[{{Wright}(1970)}]{1970VA.....12..147W}
{Wright}, K.~O. 1970, Vistas in Astronomy, 12, 147

\bibitem[{{Zuckerman} {et~al.}(1995){Zuckerman}, {Kim}, \&
  {Liu}}]{1995ApJ...446L..79Z}
{Zuckerman}, B., {Kim}, S.~S., \& {Liu}, T. 1995, \apjl, 446, L79

\end{thebibliography}

\end{document}